\documentclass[twocolumn,preprintnumbers]{revtex4}
\usepackage{epsf}

\begin{document}
\preprint{KUNS-2478}
\title{Gamov-Teller transitions from $^{14}$N ground 
to  $^{14}$C ground and excited states}
\author{Yoshiko Kanada-En'yo}
\affiliation{Department of Physics, Kyoto University, Kyoto 606-8502, Japan}
\author{Tadahiro Suhara}
\affiliation{Matsue College of Technology, Matsue 690-8518, Japan}
\begin{abstract}
Gamov-Teller transitions from the $^{14}$N ground state to the 
$^{14}$C ground and excited states were investigated, 
based on the model of antisymmetrized molecular dynamics.
The calculated strengths for the allowed transitions to the
$0^+$, $1^+$, and $2^+$ states of $^{14}$C were compared with 
the experimental data measured by high-resolution
charge-exchange reactions.
The calculated GT transition to the $2^+_1$ state is strong while  
those to 
the $0^+_{2,3}$ and $2^+_{2,3}$ states having 
dominant $2\hbar\omega$ excited configurations are relatively weak.
The present calculation can not describe the anonymously long life time of $^{14}$C, though the strength of the $^{14}$C ground state 
is somewhat suppressed because of the cluster (many-body) correlation
in the ground states of $^{14}$C and $^{14}$N.
\end{abstract}

\maketitle

\section{Introduction}
In light-mass nuclei, cluster structures often appear in ground and excited states. 
Spatially developed cluster structures are found in excited states of stable and unstable nuclei, while 
cluster components are 
also contained in the ground states. 
The cluster component usually contains 
many-body correlation, and 
in terms of the spherical shell model, it is expressed by 
mixing of higher-shell configurations beyond the simple $0\hbar\omega$ configuration. We call the correlation 
in the ground state caused by the cluster component 
"ground-state cluster correlation".

One of the typical examples of the 
ground-state cluster correlation is the $3\alpha$ cluster structure 
in $^{12}$C; 
such cluster structures develop
remarkably in excited states of $^{12}$C. Also in the ground state, the $3\alpha$ cluster component is 
significantly mixed into the $p_{3/2}$-shell closed configuration,
which is the lowest state in the uncorrelated $j$-$j$ 
coupling state. 
As a result of the mixing of the cluster component, 
the ground band of $^{12}$C exhibits oblate deformation.
This deformation can be easily understood by 
$3\alpha$ cluster models, while it is difficult to describe such a
deformation with many mean-field calculations such as 
the Hartree-Fock calculation showing the spherical 
ground state of $^{12}$C \cite{Sagawa:2004ut}.

Also in $^{14}$C, various cluster structures such as $^{10}$Be+$\alpha$ and $3\alpha+2n$ 
have been suggested in excited states \cite{Soic:2003yg,Price:2007mm,oertzen04,Itagaki:2004zz,Suhara:2010ww}. 
In our previous work  \cite{Suhara:2010ww}, which employed a method of antisymmetrized molecular dynamics (AMD), 
we discussed not only the cluster structures in excited states 
but also the cluster component in low-lying states and showed that 
the ground and low-lying states contain the cluster component (the cluster correlation) 
even though $^{14}$C is a neutron $p$-shell closed nucleus.

Recently, the Gamov-Teller (GT) transition strengths for excited states of light nuclei have been extensively studied by use of
experiments on high resolution charge-exchange reactions
\cite{Fujita:2005zz}. 
The observed $B(GT)$ values can be useful information to clarify the structure of excited states.
For $A=14$ nuclei, 
the GT strength distributions for excited 
states of $^{14}$C and the mirror $^{14}$O were studied 
in charge-exchange reactions on $^{14}$N \cite{Negret:2006zz}.
The measured $B(GT)$ distributions to $0^+$, $1^+$, and $2^+$ states 
up to the excitation energy $E_x=15$ MeV suggest the predominant strengths of $2^+$ states. In comparison with the large-scale no-core 
shell-model (NCSM) calculation \cite{Aroua:2002xm}, 
it was shown that the NCSM calculation does not reproduce 
the detailed GT strengths of excited $0^+$ and $2^+$ states, and the possible need 
to include cluster structure in these light nuclei 
was suggested \cite{Negret:2006zz}. 

As known from the anomalous long life time of $^{14}$C,
the strong suppression of the GT transition to the ground $^{14}$C
is another issue to be solved. The GT strength for the ground-ground transition 
from $^{14}$N to $^{14}$C is several orders smaller than the simple shell-model calculation without fine tuning of interaction 
\cite{inglis53,jancovici54,Visscher,Fayache:1999xs,Aroua:2002xm}.
It was suggested that the GT matrix element can vanish 
because of the accidental cancellation of the matrix element in the
$p$-shell configurations \cite{jancovici54}; the vanishing was demonstrated 
by tuning the spin-orbit and tensor forces, and also shown recently 
by adjusting the three-body terms in chiral effective field theory
in both the conventional shell model and
large-scale shell model calculations \cite{jancovici54,Visscher,Fayache:1999xs,Aroua:2002xm,Holt:2007ih,Holt:2009ty,Maris:2011as}.

In spite of sophisticated works with NCSM focusing on the GT strength 
for the ground-ground transition, the GT transitions to excited states of $^{14}$C
have not been well investigated. The large-scale NCSM calculation 
including $6\hbar\omega$ model space for excited states 
was performed by Aroua {\it et al.} \cite{Aroua:2002xm}, 
and it suggests that the inclusion of higher-shell configuration  
has significant effects on the GT strength for excited states 
as well as the ground state of $^{14}$C.
However, the calculation neither reproduces 
the experimental spectra
of excited $0^+$ and $2^+$ states in the $E_x=6\sim 10$ MeV region nor 
describes the GT strength distributions. 

In this paper, we study the GT transitions from the $1^+$ ground state of 
$^{14}$N to the ground and excited $0^+$, $1^+$, 
and $2^+$ states of $^{14}$C 
based on a method of AMD
\cite{ENYObc,AMDsupp-rev}. The AMD method 
has proven useful for describing cluster states 
as well as shell-model states, and it is suitable for investigating 
cluster structures in excited states as well as
the ground-state cluster correlation.
For the study of the ground and excited states of $^{14}$C,
we perform the variation after total-angular-momentum (spin) and parity projections in the AMD 
framework (AMD+VAP) \cite{KanadaEn'yo:1998rf}.
We obtain the $0^+_1$, $1^+_1$ and $2^+_1$
states with dominant $0\hbar\omega$ components and significant 
mixing of higher-shell components coming from the cluster components.
We also obtain the $0^+_{2,3}$ and $2^+_{2,3}$ states having developed cluster structures containing dominant 
$2\hbar\omega$ components. 
The calculated energy spectra and GT strengths are compared with the 
experimental data and also with the large-scale NCSM calculation including 
$6\hbar\omega$ configurations. 
We also apply the generator coordinate method (GCM) \cite{GCM} 
to the AMD model 
with the constraint (constraint AMD+GCM) of the deformation parameters 
$\beta$ and $\gamma$ as done in the previous work \cite{Suhara:2010ww,Suhara:2009jb}. The 
contributions of the cluster correlation to the GT transitions to the $0^+_1$ and $2^+_1$ states are discussed. 

In the present work, we use a phenomenological 
effective nuclear interaction consisting of 
central and spin-orbit forces. The adopted central interaction is
the modified Volkov force \cite{MVOLKOV} supplemented by the finite-range spin-orbit force,
which successfully reproduces the energy spectra of $^{12}$C in the 
AMD+VAP calculation \cite{KanadaEn'yo:1998rf,KanadaEn'yo:2006ze}.
With such a simple effective interaction used in the present work, 
it is difficult to describe the vanishing of the GT transition 
to the $^{14}$C ground state.
For the ground state, we only show that the GT strength can be somewhat reduced by the ground-state cluster correlation
but do not discuss the origin of the vanishing (the several-order reduction) of the GT strength, 
which may be caused by the accidental cancellation of GT matrix elements. 

This paper is organized as follows. 
In the subsequent section, the formulation and Hamiltonian of 
the present calculation are explained.
The results are shown 
in Section \ref{sec:results}. 
In Section \ref{sec:discussions}, structures of the ground and excited states are
discussed while focusing on the cluster correlation. 
The effect of the cluster correlation on the GT strengths 
is also discussed on the basis of the $\beta$-$\gamma$ constraint 
AMD calculation.
Finally, a summary is given in Section \ref{sec:summary}.

\section{Formulation}\label{sec:formulation}
To describe $^{14}$C and $^{14}$N,
we apply the AMD+VAP method.
We also apply the $\beta$-$\gamma$ constraint AMD+GCM method
and obtain results similar to those for AMD+VAP result.

The AMD+VAP method is the same one used for the 
study of $^{12}$C in Refs.~\cite{KanadaEn'yo:1998rf,KanadaEn'yo:2006ze}, and the $\beta$-$\gamma$ constraint AMD+GCM is basically 
the same as the method 
used in the previous work for $^{14}$C \cite{Suhara:2010ww}.
For the details of these frameworks, the readers are referred to 
Refs.~\cite{Suhara:2010ww,AMDsupp-rev,KanadaEn'yo:1998rf,KanadaEn'yo:2006ze,Suhara:2009jb}.

\subsection{AMD wave functions}

In the AMD method, a basis wave function of an $A$-nucleon system 
is described by a Slater determinant of single-particle Gaussian wave packets,
\begin{equation}
 \Phi_{\rm AMD}({\bf Z}) = \frac{1}{\sqrt{A!}} {\cal{A}} \{
  \varphi_1,\varphi_2,...,\varphi_A \}.
\end{equation}
The $i$th single-particle wave function $\varphi_i$ is written as a product of
spatial, intrinsic spin, and isospin
wave functions:
\begin{eqnarray}
 \varphi_i&=& \phi_{{\bf X}_i}\chi_i\tau_i,\\
 \phi_{{\bf X}_i}({\bf r}_j) & = &  \left(\frac{2\nu}{\pi}\right)^{4/3}
\exp\bigl\{-\nu({\bf r}_j-\frac{{\bf X}_i}{\sqrt{\nu}})^2\bigr\},
\label{eq:spatial}\\
 \chi_i &=& (\frac{1}{2}+\xi_i)\chi_{\uparrow}
 + (\frac{1}{2}-\xi_i)\chi_{\downarrow}.
\end{eqnarray}
$\phi_{{\bf X}_i}$ and $\chi_i$ are the spatial and spin functions, and 
$\tau_i$ is the isospin
function fixed either up (proton) or down (neutron). 
The width parameter $\nu$ is fixed at the same value
$\nu=0.19$ fm$^{-2}$ as that used in the study
of $^{12}$C \cite{KanadaEn'yo:2006ze}.
Accordingly, an AMD wave function
is expressed by a set of variational parameters
${\bf Z}\equiv 
\{{\bf X}_1,{\bf X}_2,\cdots, {\bf X}_A,\xi_1,\xi_2,\cdots,\xi_A \}$
which express Ganssian center positions and spin orientations of $A$ nucleons.

\subsection{AMD+VAP method}
In the AMD+VAP method, 
the energy variation is performed after the spin and parity projections 
in the AMD model as done in previous work on $^{12}$C 
\cite{KanadaEn'yo:1998rf,KanadaEn'yo:2006ze}. 
For the lowest $J^\pi$ state,
the parameters ${\bf X}_i$ and $\xi_{i}$($i=1\sim A$) are varied to
minimize the energy expectation value of the Hamiltonian,
$\langle \Phi|H|\Phi\rangle/\langle \Phi|\Phi\rangle$,
with respect to the spin-parity eigen wave function projected 
from an AMD wave function; $\Phi=P^{J\pi}_{MK}\Phi_{\rm AMD}({\bf Z})$.
Here, $P^{J\pi}_{MK}$ is the spin-parity projection operator.
After the energy variation by
a frictional cooling method \cite{AMDsupp-rev},
the optimum AMD wave function
$\Phi_{\rm AMD}({\bf Z}^{J^\pi})$
is obtained. The obtained wave function $\Phi_{\rm AMD}({\bf Z}^{J^\pi})$
approximately describes the intrinsic wave function for 
the lowest $J^\pi$ state. 
For higher $J^\pi_k$ ($k\ge 2$) states, the energy variation after the spin and parity 
projections is performed for the component orthogonal to the lower 
$J^\pi$ states. 
Then, for each $J^\pi_k$, the optimum parameter solution ${\bf Z}^{J^\pi_k}$ is obtained. In the case of $^{14}$C, 
we perform the VAP calculations for the 
$0^+_{1,2,3}$, $1^+_1$, and $2^+_{1,2,3}$ states and 
obtain seven sets of  parameters ${\bf Z}$. After the VAP procedure, 
final wave functions are 
calculated by superposing the spin-parity eigen wave functions 
projected from these seven AMD wave functions $\Phi_{\rm AMD}({\bf Z}^{\alpha})$
obtained by the VAP.
Here $\alpha$ is the label for seven VAP states,
$\alpha$=$0^+_{1,2,3}$, $1^+_1$, and $2^+_{1,2,3}$ states.
Namely, the final wave functions for the $J^\pi_n$ states 
are expressed as 
\begin{equation}\label{eq:diago}
|\Psi^{^{14}{\rm C}}_{\rm VAP}(J^\pi_n)\rangle=\sum_{K,\alpha} c^{J^\pi}_n(K,\alpha) 
|P^{J\pi}_{MK}\Phi_{\rm AMD}({\bf Z}^{\alpha})\rangle,
\end{equation}
where the coefficients $c^{J^\pi}_n(K,\alpha)$ are determined by
diagonalization of norm and Hamiltonian matrices.
For the ground state of $^{14}$N, the VAP is performed 
for $J^\pi=1^+$. Then, the $^{14}$N ground state 
wave function is described by the spin-parity eigen state projected 
from the $\Phi_{\rm AMD}({\bf Z}^{1^+_1})$ with $K$-mixing.

\subsection{$\beta$-$\gamma$ constraint AMD+GCM}\label{sec:amd+gcm}
In the $\beta$-$\gamma$ constraint AMD+GCM method, 
the energy variation is performed after the parity projection 
but before the spin projection under certain constraints.
Namely, we  perform the energy variation for the parity projected wave function,
$\Phi=P^{\pi}\Phi_{\rm AMD}({\bf Z})$ with the constraint on the quadrupole 
deformation parameters $\beta$ and $\gamma$. Here, $P^\pi$ is the parity projection operator.
The deformation parameters $\beta$ and $\gamma$ are defined as 
\begin{eqnarray}
	&\beta \cos \gamma \equiv \frac{\sqrt{5\pi}}{3} 
		\frac{2\langle {z}^{2} \rangle -\langle {x}^{2} \rangle 
   -\langle {y}^{2} \rangle }{R^{2}}, \\
	&\beta \sin \gamma \equiv \sqrt{\frac{5\pi}{3}} 
		\frac{\langle {x}^{2} \rangle -\langle {y}^{2} \rangle }{R^{2}}, \\
&R^{2} \equiv \frac{5}{3} \left( \langle {x}^{2} \rangle + \langle {y}^{2} \rangle 
		+ \langle {z}^{2} \rangle \right).
\end{eqnarray}
For a given set of constraint parameters ($\beta_i$,$\gamma_i$),  
we impose the constraints: $\beta \cos \gamma= \beta_i\cos\gamma_i$, 
$\beta \sin \gamma= \beta_i\sin\gamma_i$, and 
\begin{equation}
 \frac{\langle {xy} \rangle}{R^2}=\frac{\langle {yz} \rangle}{R^2}
     =\frac{\langle {zx} \rangle}{R^2}=0.
\end{equation}
After the energy variation with the constraints, we obtain the optimized wave function
$\Phi=P^{\pi}\Phi_{\rm AMD}({\bf Z}^{(\beta_i,\gamma_i)})$
for the $i$th set of deformation 
parameters $(\beta_i,\gamma_i)$.
For the constraint parameters $(\beta_i,\gamma_i)$, 
we take the triangular lattice points with 
mesh size 0.05 on the $\beta$-$\gamma$ plane 
as done in Refs.~\cite{Suhara:2009jb,Suhara:2010ww}.
We truncate the $(\beta\cos \gamma, \beta\sin \gamma)$ region as $\beta_i \sin \gamma_i \le -(\beta_i \cos \gamma_i -1 )/2$ and $\beta_i \sin \gamma_i \le -(\beta_i \cos \gamma_i -0.75 )/2$ 
and use a total of 121 and 72 mesh points for 
$^{14}$C and $^{14}$N, respectively,
to save numerical cost. These truncations 
do not affect the results of low-lying states.

To obtain the wave functions for $J^\pi_k$ states, 
we superpose the spin-parity projected AMD
wave functions $P^{J\pi}_{MK}\Phi_{\rm AMD}
({\bf Z}^{(\beta_i,\gamma_i)})$
using GCM \cite{GCM}.
Then the final wave functions for the $J^\pi_n$ states are described as 
\begin{equation}\label{eq:diago}
|\Psi^{^{14}{\rm C}}_{\beta\gamma\mathchar`-{\rm MC}}(J^\pi_n)\rangle
=\sum_{i,K} c^{J^\pi}_n(K,\beta_i,\gamma_i) 
|P^{J\pi}_{MK}\Phi_{\rm AMD}({\bf Z}^{(\beta_i,\gamma_i)})\rangle,
\end{equation}
where the coefficients $c^{J^\pi}_n(K,\beta_i,\gamma_i)$ are determined 
by solving the Hill-Wheeler equation, i.e., the 
diagonalization of the norm and Hamiltonian matrices. 
The final wave function 
$\Psi^{^{14}{\rm C}}_{\beta\gamma\mathchar`-{\rm MC}}(J^\pi_n)$ is the multiconfiguration (MC) state, which is expressed by the 
linear combination of various configurations 
$P^{J\pi}_{MK}\Phi_{\rm AMD}({\bf Z}^{(\beta_i,\gamma_i)})$
 obtained by the $\beta$-$\gamma$ constraint AMD.

\subsection{Effective interactions}
We use the same effective nuclear interaction as that 
used in the previous calculation of 
$^{12}$C \cite{KanadaEn'yo:2006ze}.
It is the MV1 force \cite{MVOLKOV} 
 for the central force 
supplemented by the two-body spin-orbit force given by 
the two-range Gaussian form with the range parameters being the same 
as those of the G3RS force \cite{LS}.
The Coulomb force is approximated by using a seven-range
Gaussian form. 
The spin-orbit strengths are taken to be $u_{I}=-u_{II}=3000$ MeV.
The Majorana, Bartlett, 
and Heisenberg parameters in the MV1 force are taken to be
(A) $m=0.62$, and $b=h=0.125$ as well as the 
parameters (B) $m=0.62$ and $b=h=0$ used in Ref.~\cite{KanadaEn'yo:2006ze}.
We also present the results with a different Majorana parameter, 
(C) $m=0.58$ and $b=h=0.125$, and (D) $m=0.58$ and $b=h=0$ to show that the interaction parameter dependence is minor.

In the present work, the MV1 force is adopted as the 
effective central interaction instead of the Volkov force 
used in the previous work on $^{14}$C \cite{Suhara:2010ww}. 
In Ref.~\cite{Suhara:2010ww},  
the excitation energies calculated with the $\beta$-$\gamma$
constraint AMD using the Volkov No.2 force 
largely overestimate the experimental excitation energies of $^{14}$C.
This may come from the overbinding problem of the Volkov force for heavier
mass nuclei.
The MV1 force is the interaction modified from the Volkov force,
and it consists of the finite-range two-body force and 
the zero-range three-body term. The force reasonably reproduces 
the energy spectra of $p$-shell and $sd$-shell nuclei.

\section{Results}\label{sec:results}

\begin{table}[ht]
\caption{
\label{tab:ex-gt}
Excitation energies and $B(E2)$ of 
$^{14}$C, and $B(GT)$ from $^{14}$N$(1^+_1)$.
The theoretical values calculated with AMD+VAP using the
interactions from 
sets A, B, C, and D  are shown as well as those calculated with the 
$\beta$-$\gamma$ constraint AMD+GCM using the set A interaction.
The experimental data are taken from Refs.~\cite{Negret:2006zz,AjzenbergSelove:1991zz}.
}
\begin{center}
\begin{tabular}{c|cc|cc}
\hline
&  \multicolumn{2}{c}{AMD+VAP(A)} &	 \multicolumn{2}{c}{AMD+VAP(B)} \\	
	&	$E_x$ (MeV)	&	$B(GT)$	&	$E_x$ (MeV)	&	$B(GT)$	\\
$^{14}$C$(0^+_1)$	&	0	&	0.07 	&	0	&	0.09 	\\
$^{14}$C$(0^+_2)$	&	10.3 	&	0.001 	&	10.5 	&	0.003 	\\
$^{14}$C$(0^+_3)$	&	16.0 	&	0.00002 	&	16.8 	&	0.00003 	\\
$^{14}$C$(2^+_1)$	&	7.9 	&	2.4 	&	7.5 	&	2.5 	\\
$^{14}$C$(2^+_2)$	&	11.2 	&	0.24 	&	11.5 	&	0.17 	\\
$^{14}$C$(2^+_3)$	&	14.9 	&	0.03 	&	15.3 	&	0.03 	\\
$^{14}$C$(1^+_1)$	&	12.2 	&	0.21 	&	12.8 	&	0.22 	\\
	&		\multicolumn{2}{c}{$B(E2)$ (e$^2$fm$^4$)} 
&\multicolumn{2}{c}{$B(E2)$ (e$^2$fm$^4$)} 	\\
$^{14}$C;$2^+_1\rightarrow 0^+_1$ 	
&	\multicolumn{2}{c}{6.6}	&\multicolumn{2}{c}{7.3}		\\
\hline									
&  \multicolumn{2}{c}{AMD+VAP(C)} &	 \multicolumn{2}{c}{AMD+VAP(D)} \\		
	&	$E_x$ (MeV)	&	$B(GT)$	&	$E_x$ (MeV)	&	$B(GT)$	\\
$^{14}$C$(0^+_1)$	&	0	&	0.12 	&	0	&	0.13 	\\
$^{14}$C$(0^+_2)$	&	10.9 	&	0.001 	&	10.8 	&	0.002 	\\
$^{14}$C$(0^+_3)$	&	16.5 	&	0.0001 	&	17.7 	&	0.0001 	\\
$^{14}$C$(2^+_1)$	&	10.1 	&	2.1 	&	9.6 	&	2.4 	\\
$^{14}$C$(2^+_2)$	&	11.7 	&	0.75 	&	11.9 	&	0.42 	\\
$^{14}$C$(2^+_3)$	&	15.2 	&	0.002 	&	16.4 	&	0.003 	\\
$^{14}$C$(1^+_1)$	&	14.5 	&	0.31 	&	15.1 	&	0.32 	\\
	&		\multicolumn{2}{c}{$B(E2)$ (e$^2$fm$^4$)} 
&\multicolumn{2}{c}{$B(E2)$ (e$^2$fm$^4$)} 	\\
$^{14}$C;$2^+_1\rightarrow 0^+_1$ 	
&	\multicolumn{2}{c}{4.1}	&\multicolumn{2}{c}{4.9}		\\
\hline									
&  \multicolumn{2}{c}{$\beta$-$\gamma$-AMD+GCM(A)} & \multicolumn{2}{c}{exp.}\\					
	&	$E_x$ (MeV)	&	$B(GT)$	&$E_x$ (MeV)	&	$B(GT)$			\\
$^{14}$C$(0^+_1)$	&	0	&	0.16 	&		0	&	1.90E-06		\\
$^{14}$C$(0^+_2)$	&	9.0 	&	0.0002 	&	6.589	&	0.056		\\
$^{14}$C$(0^+_3)$	&	14.2 	&	0.00005 	&		&		\\
$^{14}$C$(2^+_1)$	&	7.2 	&	2.3 	&	7.012	&	0.45	\\
$^{14}$C$(2^+_2)$	&	9.7 	&	0.33 	&	8.318	&	0.37	\\
$^{14}$C$(2^+_3)$	&	12.2 & 0.10 & 10.425	&	0.098		\\
$^{14}$C$(2^+_4)$	&	14.4 & 0.04 & 	&		\\
$^{14}$C$(1^+_1)$	&	12.9 	&	0.25 	&	11.306	&	0.072		\\
	&		\multicolumn{2}{c}{$B(E2)$ (e$^2$fm$^4$)} 
&\multicolumn{2}{c}{$B(E2)$ (e$^2$fm$^4$)} 	\\
$^{14}$C;$2^+_1\rightarrow 0^+_1$ 	&	\multicolumn{2}{c}{7.4}	&	\multicolumn{2}{c}{3.74} \\
\end{tabular}
\end{center}
\end{table}

With the AMD+VAP method we calculate 
the $1^+$ ground state of $^{14}$N. The magnetic moment and
electric quadrupole moment calculated with AMD+VAP using 
the set A interaction are $\mu=+0.34$ $(\mu_N)$ and $Q=+0.92$ (e$\cdot$fm$^2$). They reasonably agree with the experimental values,
$\mu_{\rm exp}=+0.40376100(6)$ $(\mu_N)$ and 
$Q=+1.93(8)$ (e$\cdot$fm$^2$).
We also apply the AMD+VAP method to
the $0^+_{1,2,3}$, $1^+_1$, and $2^+_{1,2,3}$ states of
$^{14}$C and calculate 
the $GT$ transition 
strengths from the $^{14}$N ground state to $^{14}$C states.
The GT transition strength $B(GT)$ is given as
\begin{equation}
B(GT)=\left(\frac{g_A}{g_{V}}\right)^2 \frac{1}{2J_i+1}|\langle J_f||\sigma\tau_\pm||J_i\rangle|^2.
\end{equation}
Here $g_A/g_V=1.251$ is the ratio of the GT to Fermi coupling constant.
In principle, the GT transition from the $J^\pi=1^+$, $T=0$ state 
is allowed for $J^\pi=0^+$, $1^+$, and $2^+$, $T=1$ states.

The results of $B(GT)$ for the transitions from 
$^{14}$N$(1^+_1)$ to $^{14}$C($0^+_{1,2,3}$, $1^+_1$, 
and $2^+_{1,2,3}$) as well as those of 
the excitation energies and $B(E2)$ of $^{14}$C are shown 
in Table \ref{tab:ex-gt}, compared with the experimental data.
The theoretical values obtained by the AMD+VAP calculation using 
four sets of interaction parameters are listed. 
Properties of the ground and excited states of $^{14}$C are not strongly dependent on the interaction parameters within the present calculation.
The calculated results obtained the 
$\beta$-$\gamma$ constraint AMD+GCM using the set A interaction 
are also shown in the table.
They are qualitatively 
consistent with those obtained with the AMD+VAP calculations.

The calculated $B(GT)$ for $^{14}$C($0^+_1$) is 
relatively small compared with those for $J^\pi=1^+_1$ 
and $2^+_1$ states. However,  
the present calculations fail to describe 
the vanishing of the GT strength known from the 
anomalously long life time of $^{14}$C.
For $^{14}$C($2^+_1$), the GT strength is $B(GT)=2 \sim 3$ in the present calculation 
and is larger than the experimental value $B(GT)=0.45$ measured by charge-exchange reactions. 
We obtain the $0^+_{2}$ and $2^+_{2}$ states with dominant 
$2\hbar\omega$ neutron excited configurations
around $E_x=10$ MeV. 
The features of these two states,
such as the $0^+$-$2^+$ level spacing and the GT transition, reasonably agree with those of 
the experimental $0^+_2$ and $2^+_2$ states; therefore, 
we assign them to the experimental
$^{14}$C($0^+$,6.6 MeV) and $^{14}$C($2^+$,8.3 MeV).
We also obtain the $0^+_{3}$ and $2^+_{3}$ states dominated by 
$2\hbar\omega$ neutron excited configurations around $E_x=15$ MeV.

\begin{figure}[th]
\epsfxsize=0.35\textwidth
\centerline{\epsffile{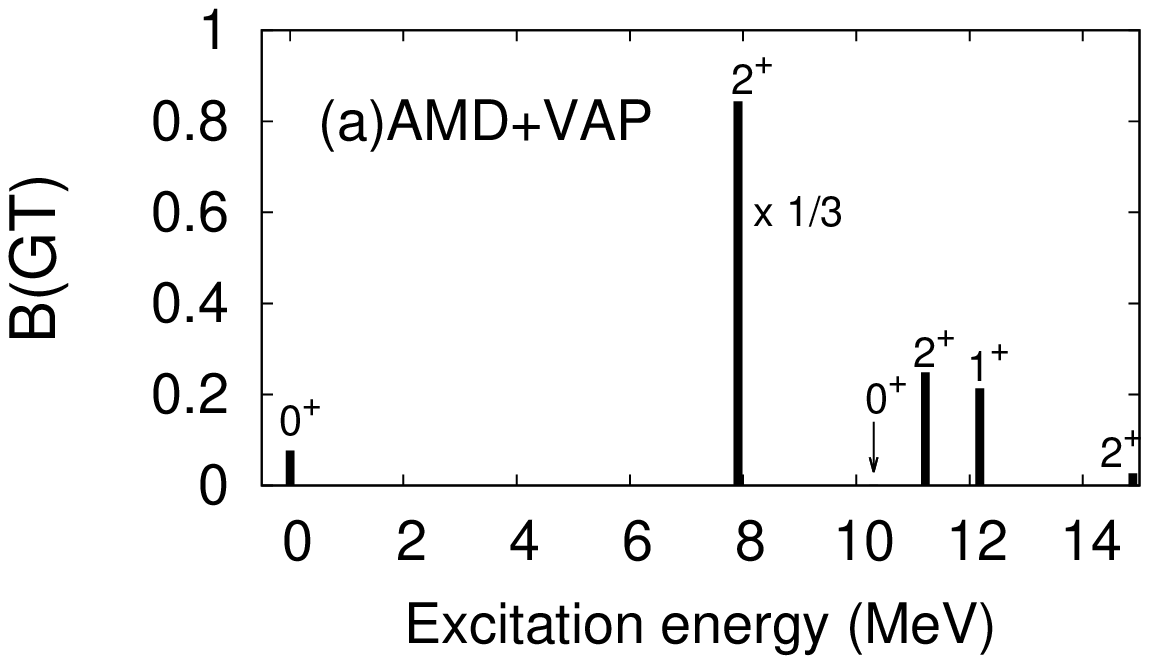}}
\epsfxsize=0.35\textwidth
\centerline{\epsffile{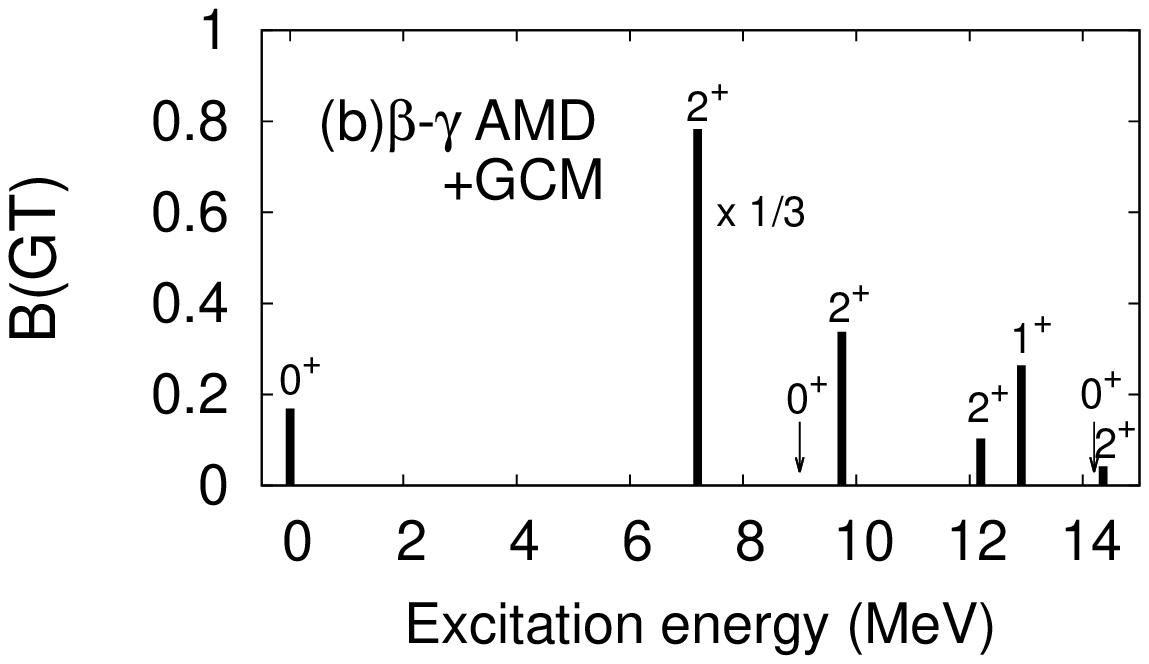}}
\epsfxsize=0.35\textwidth
\centerline{\epsffile{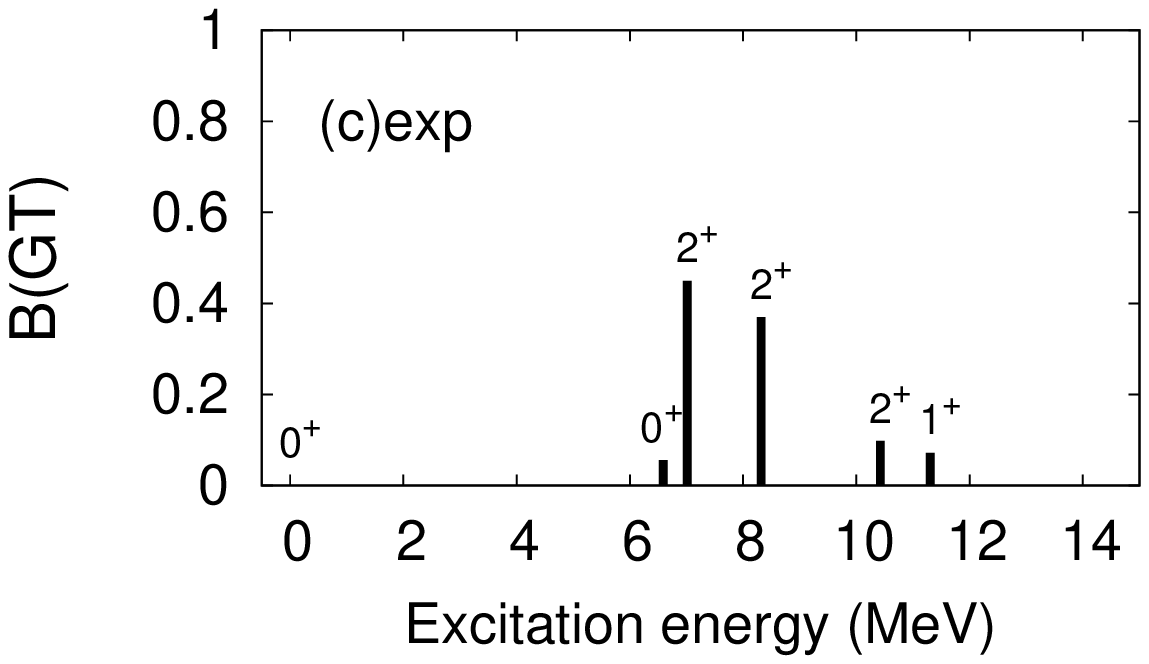}}
\epsfxsize=0.35\textwidth
\centerline{\epsffile{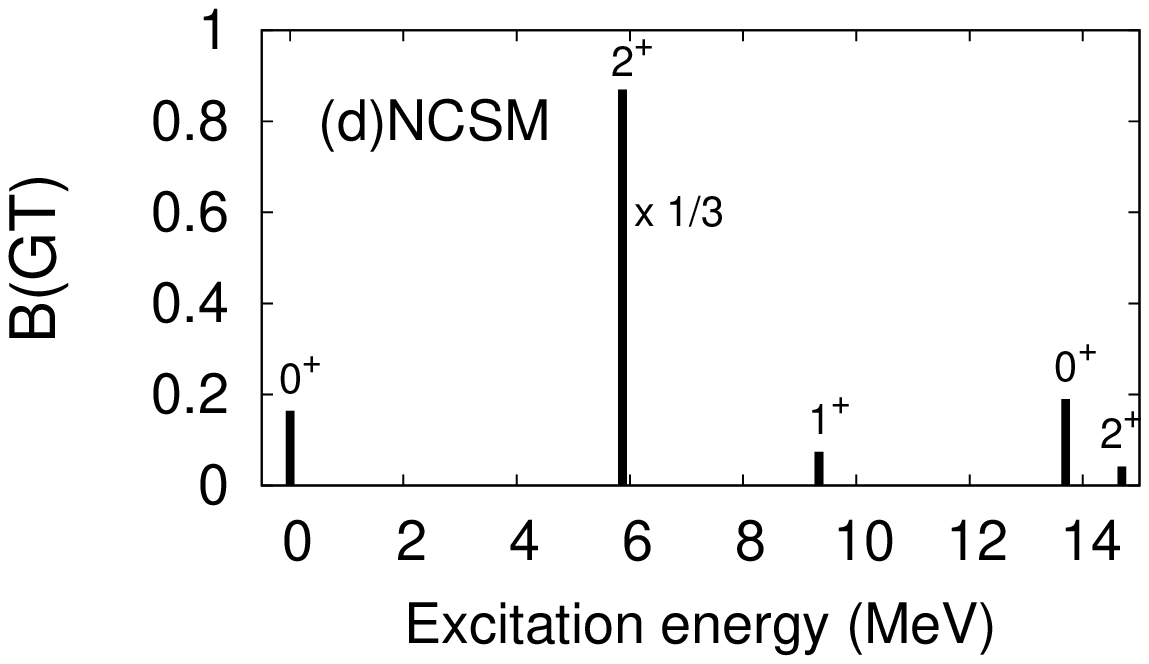}}
\caption{\label{fig:bgt} 
The $B(GT)$ distributions for transitions from the $1^+$ ground state of $^{14}$N to 
$J^\pi=0^+$,$1^+$, and $2^+$ states of $^{14}$C.
(a) The theoretical values calculated with the AMD+VAP using
the set A interaction, (b) those calculated with the 
$\beta$-$\gamma$ constraint AMD+GCM, 
(c) the experimental data taken from Refs.~\cite{Negret:2006zz,AjzenbergSelove:1991zz}, and 
(d) the theoretical results from Ref.~\cite{Aroua:2002xm} of the NCSM calculation
with the $6\hbar\omega$ model space using the effective interactions derived from
Argonne V8' interaction.
}
\end{figure}

The calculated $B(GT)$ distributions for 
$^{14}{\rm N}(1^+_1)\rightarrow ^{14}{\rm C}(0^+,1^+,2^+)$ obtained 
with the AMD+VAP and $\beta$-$\gamma$ constraint AMD+GCM using 
the set A interaction 
are compared with the experimental data and 
the large-scale $6\hbar\omega$ NCSM calculation with AV8'
interaction \cite{Aroua:2002xm}
in Fig.~\ref{fig:bgt}. 
Qualitative features of the $B(GT)$ distributions obtained 
in the present calculations
are in reasonable agreement with the experimental 
$B(GT)$ distributions except for $B(GT)$ for $^{14}{\rm C}(2^+_1)$. 
The calculated $B(GT)$ for the $2^+_1$ state is largest 
and is as much as that of the NCSM calculation. 
In the experimental $B(GT)$ distribution for $2^+$ states, 
relatively larger populations were observed compared with $0^+$ and $1^+$ states. 
However, the calculations overestimate 
the absolute value of the experimental $B(GT)$ 
for the $2^+_1$ state.
This might indicate that the description of the final state ($^{14}{\rm C}(2^+_1)$)
and/or the initial state ($^{14}$N$(1^+_1)$) is not sufficient 
in the present calculations.

Compared with the NCSM calculation, the present 
$B(GT)$ for $^{14}{\rm C}(0^+_1)$ is the same order as 
that of the large-scale $6\hbar\omega$ NCSM calculation using the 
effective interaction derived from the AV8' interaction.
It should be noted that the vanishing of the GT matrix 
element for $^{14}{\rm C}(0^+_1)$ has been discussed in several 
NCSM studies
by fine tuning of the 
interactions \cite{Fayache:1999xs,Aroua:2002xm,Holt:2007ih,Holt:2009ty,Maris:2011as}.
For $0^+_2$ and $2^+_2$ states, the correspondence to the
experimental $B(GT)$ distributions seems better in the present result
than in the NCSM calculation. In the NCSM calculation, 
the corresponding states might be missing or their excitation energies 
might be overestimated. In the present results,  
$0^+_2$ and $2^+_2$ states contain cluster correlations 
resulting in the mixing of higher-shell components 
of proton and neutron excitations. Usually, shell model calculations 
are not suitable for describing such cluster states.
 
\section{Discussion}\label{sec:discussions}
In the present result, even the ground states of 
$^{14}$C and $^{14}$N have significant mixing 
of proton and neutron excitations 
from the lowest $0\hbar\omega$ configuration
because of the cluster correlations. 
In this section, we discuss intrinsic structures and cluster correlations
in $^{14}$C and $^{14}$N. 
To show the mixing of excited configurations, we analyze
the probability of higher-shell components.
The effect of the cluster correlations on the GT strengths for $^{14}$C$(0^+_1)$ and 
$^{14}$C$(2^+_1)$ is also discussed.

\subsection{Intrinsic structure and cluster correlation}

\begin{figure}[th]
\epsfxsize=0.48\textwidth
\centerline{\epsffile{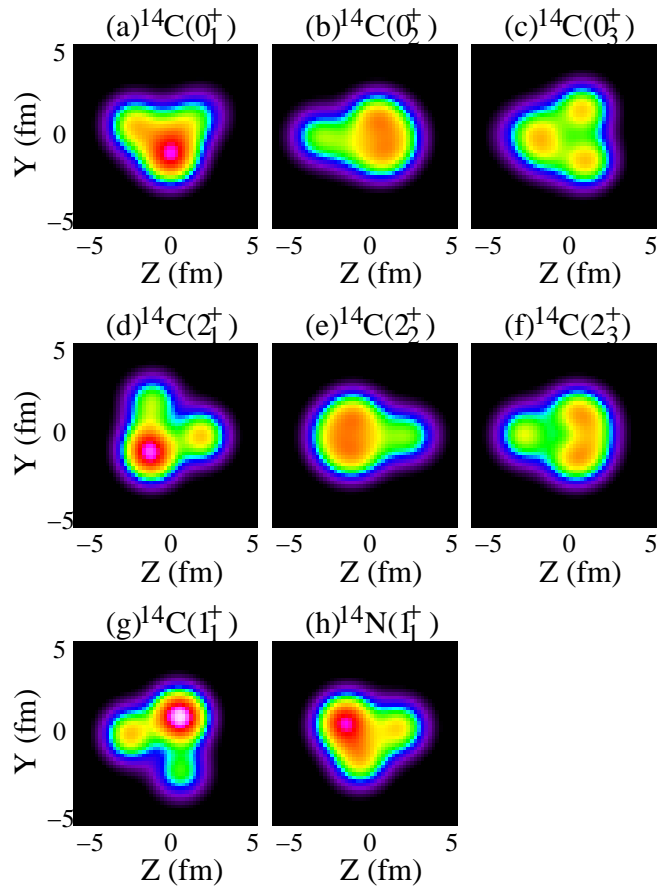}}
\caption{\label{fig:c14-vap} (Color online)
Density distribution for the intrinsic wave functions 
$\Phi_{\rm AMD}({\bf Z}^{J^\pi_k})$
of $^{14}$C and $^{14}$N obtained with the AMD+VAP
using the set A interaction. 
$x$, $y$, and  $z$ axes are chosen as 
$\langle x^2 \rangle \le \langle y^2 \rangle \le \langle z^2 \rangle$
and the matter density integrated along the
$x$ axis is plotted on the $z$-$y$ plane. 
}
\end{figure}

In the AMD+VAP calculation, the AMD wave function
$\Phi_{\rm AMD}({\bf Z}^\alpha)$ obtained by the VAP calculation 
for $\alpha=J^\pi_k$ is regarded as the intrinsic state of 
the $J^\pi_k$ state. 
As seen in the density distributions of the intrinsic wave functions 
$\Phi_{\rm AMD}({\bf Z}^\alpha)$ for $^{14}$C and $^{14}$N in 
Fig.~\ref{fig:c14-vap}, 
$\alpha$-like or $t$-like cluster correlations are found 
even in the ground states of $^{14}$C and $^{14}$N. 
In the excited states of $^{14}$C, 
further development of cluster structures is seen.

On the basis of the spherical harmonic oscillator (HO) shell model, 
those states contain significant components of excited configurations
because of the cluster correlations.
To discuss the higher-shell components beyond the lowest $0\hbar\omega$ 
configurations in each state, we calculate the occupation probability of HO quanta in the final wave function 
$\Psi^{^{14}{\rm C}}_{\rm VAP}(J^\pi_n)$ for the $J^\pi_n$ state
of $^{14}$C.
The calculations of the occupation probability are done 
following the method proposed by Suzuki {\ et al.} \cite{Suzuki:1996ui}. 
The occupation probability of a definite 
number of total HO quanta $Q$ is given by the expectation value 
$\langle P_Q \rangle$
of the following projection operator $P_Q$ to 
the eigen state of the total HO quanta operator $\sum_i a^\dagger_i a_i$,
\begin{eqnarray}
&P_Q=\frac{1}{2\pi}\int^{2\pi}_0 d\theta \exp\left[
i\theta \left(\sum^A_{i=1}a^\dagger_i a_i-Q \right) \right]\\
& a^\dagger_i =\sqrt{m\omega/2\hbar}(x_i-ip_i/m\hbar)\\ 
& a_i =\sqrt{m\omega/2\hbar}(x_i+ip_i/m\hbar),
\end{eqnarray}
where $\omega\equiv 2\hbar\nu/m$ ($\nu$ is the width parameter used in 
AMD wave functions).
The occupation probability of total proton(neutron) 
HO quanta is calculated similarly by using the isospin 
projection operator.

\begin{figure}[th]
\epsfxsize=0.48\textwidth
\centerline{\epsffile{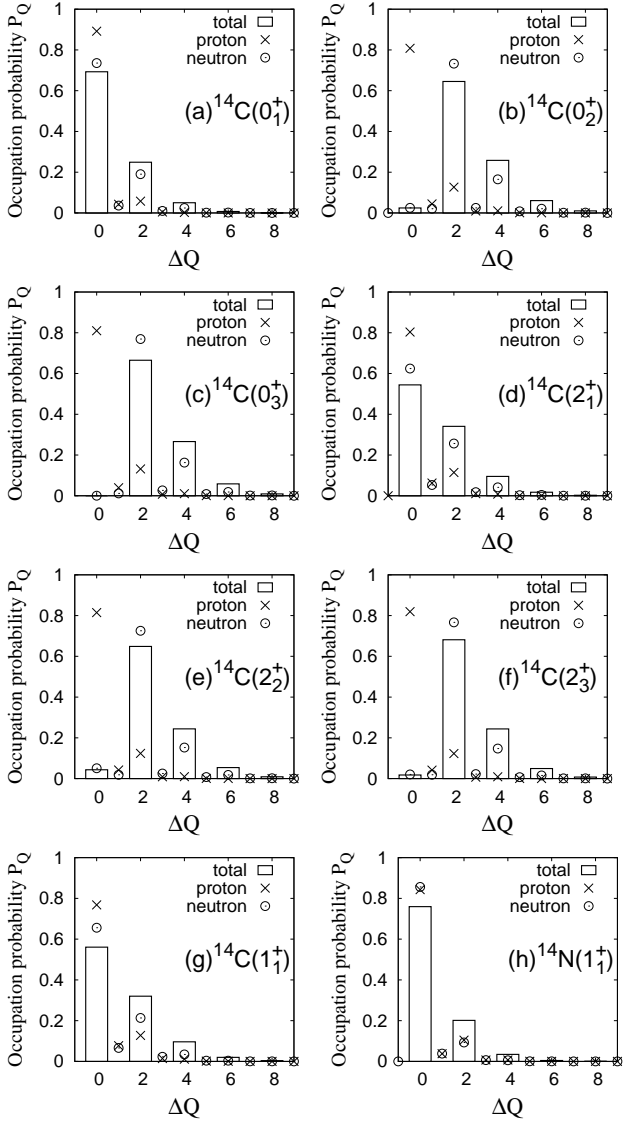}}
\caption{\label{fig:ho-quanta}
Occupation probability $\langle P_Q \rangle$
for protons, neutrons, and total nucleons
calculated with the AMD+VAP
using the set A interaction.
The calculated $\langle P_Q \rangle$ values are plotted as
functions of $\Delta Q\equiv Q-Q_{\rm min}$ 
measured from the lowest allowed HO quanta $Q_{\rm min}$.
}
\end{figure}

We calculate $\langle P_Q \rangle$
for protons, neutrons, and total nucleons and  
plot the values as functions of $\Delta Q\equiv Q-Q_{\rm min}$ 
measured from the lowest allowed HO quanta $Q_{\rm min}$. 
$\langle P_Q \rangle$ for total nucleons stands for 
the component of $\Delta Q$-$\hbar\omega$ configurations,
and that for protons (neutrons) indicates the probability
of proton (neutron) $\Delta Q$-$\hbar\omega$ excitation. 
The results of the AMD+VAP states 
$\Psi^{^{14}{\rm C}}_{\rm VAP}(J^\pi_n)$
and $\Psi^{^{14}{\rm N}}_{\rm VAP}(1^+_1)$ obtained using the 
set A interaction are shown in Fig.~\ref{fig:ho-quanta}. 

It is found that 
the ground state $^{14}$C$(0^+_1)$ is not only dominated by the $0\hbar\omega$ component 
but also contains significant components, i.e., 25\% and 5\% of 
$2\hbar\omega$ and $4\hbar\omega$ excited configurations, respectively.
In the higher-shell components, the neutron excitation is dominant and the proton excitation is minor. The mixing of proton excitation is caused by the cluster correlation. 
Also the ground state of $^{14}$N
contains 20\% of 
$2\hbar\omega$ excited configurations, a significant component.

In the excited states $^{14}$C$(2^+_1)$ and $^{14}$C$(1^+_1)$,
the $0\hbar\omega$ component is dominant but 
 is reduced to 50\% because of the  
larger probability of excited configurations than the ground state.
$^{14}$C$(0^+_{2,3})$ and $^{14}$C$(2^+_{2,3})$ have the dominant
$2\hbar\omega$ configuration with the significant mixing of 
$4\hbar\omega$ and $6\hbar\omega$ configurations.
In addition to the major neutron excitations, they also contain a
20\% component of proton excitations because of the cluster correlation

Similar features for $\langle P_Q \rangle$ are found
in the result of the $\beta$-$\gamma$ constraint AMD+GCM calculation.
Namely, significant higher-shell components are contained even 
in the ground states of $^{14}$C and $^{14}$N. 
As described in Section \ref{sec:amd+gcm}, 
the final wave function 
$\Psi^{^{14}{\rm C}}_{\beta\gamma\mathchar`-{\rm MC}}(J^\pi_n)$
in the $\beta$-$\gamma$ constraint AMD+GCM calculation
is given by the superposition of various AMD configurations on the $\beta$-$\gamma$ plane. In the framework of $\beta$-$\gamma$ constraint AMD, the spherical $\beta=0$ state corresponds to a 
$0\hbar\omega$ configuration state, while deformed states with 
finite $\beta$ and/or $\gamma$ values contain higher-shell components
in terms of the spherical HO shell model.
For $^{14}$C and $^{14}N$ systems, the finite $\beta$ 
and/or finite $\gamma$ states have cluster structure
containing more components of higher-shell configurations 
because of the cluster correlations.
This means that, in the $\beta$-$\gamma$ constraint AMD+GCM,  
higher-shell components beyond the $0\hbar\omega$ configuration are mixed in the ground-state wave function 
through the finite $\beta$ 
and/or finite $\gamma$ states in the superposition 
of basis AMD wave functions. Therefore, in this framework, 
the origin of the mixing of excited configurations can be understood 
by the deformation modes accompanied by cluster correlations.

\begin{figure}[th]
\epsfxsize=0.35\textwidth
\centerline{\epsffile{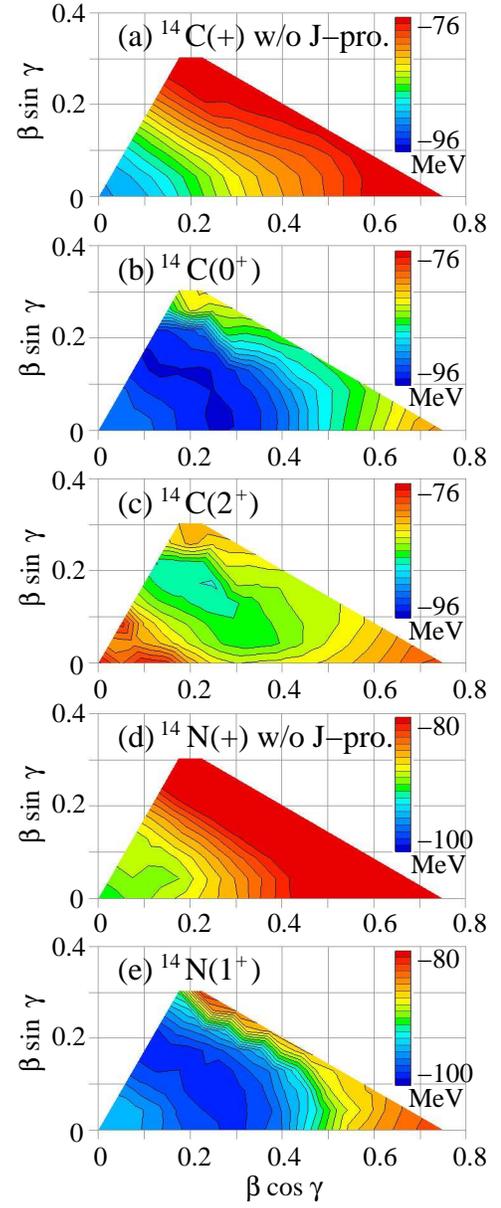}}
\caption{\label{fig:c14-bg-ene} (Color online)
Energy expectation values calculated 
with the $\beta$-$\gamma$ constraint AMD.
(a) Energy for the positive-parity state projected from 
$\Phi_{\rm AMD}({\bf Z}^{(\beta,\gamma)})$ of $^{14}$C
without the spin projection.
(b) and (c) Energy for the $0^+$ and $2^+$ states of $^{14}$C
projected from $\Phi_{\rm AMD}({\bf Z}^{(\beta,\gamma)})$.
(d) Energy for the positive-parity state projected from 
$\Phi_{\rm AMD}({\bf Z}^{(\beta,\gamma)})$ of $^{14}$N
without the spin projection.
(e) Energy for the $1^+$ state of $^{14}$N
projected from $\Phi_{\rm AMD}({\bf Z}^{(\beta,\gamma)})$.
}
\end{figure}

\begin{figure}[th]
\epsfxsize=0.48\textwidth
\centerline{\epsffile{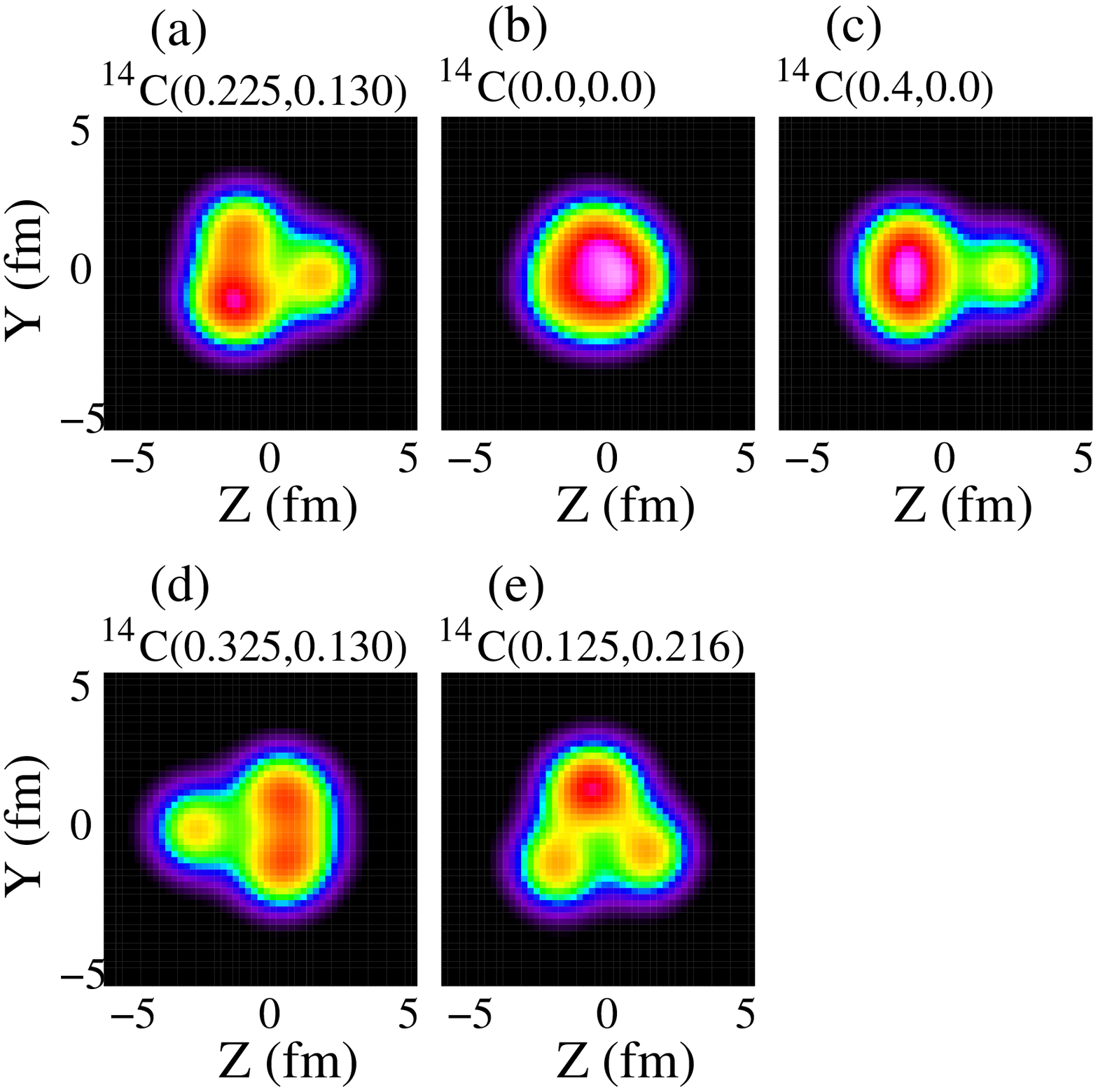}}
\caption{\label{fig:c14-bg} (Color online)
Density distribution for the intrinsic wave functions 
$\Phi_{\rm AMD}({\bf Z}^{(\beta,\gamma)})$
of $^{14}$C obtained with the $\beta$-$\gamma$ constraint AMD
using the set A interaction. 
The densities for the $J^\pi=0^+$ energy minimum 
$(\beta_{\rm min}\cos\gamma_{\rm min},
\beta_{\rm min}\sin\gamma_{\rm min})=(0.225,0.130)$ state,
and the spherical $\beta=0$, prolate 
$(\beta\cos\gamma,\beta\sin\gamma)=(0.4,0.0)$, triaxial $(0.325,0.130)$, 
and oblate $(0.125,0.216)$ states are shown.
$x$, $y$, and  $z$ axes are chosen as 
$\langle x^2 \rangle \le \langle y^2 \rangle \le \langle z^2 \rangle$
and the matter density integrated along the
$x$ axis is plotted on the $z$-$y$ plane. 
}
\end{figure}

\begin{figure}[th]
\epsfxsize=0.48\textwidth
\centerline{\epsffile{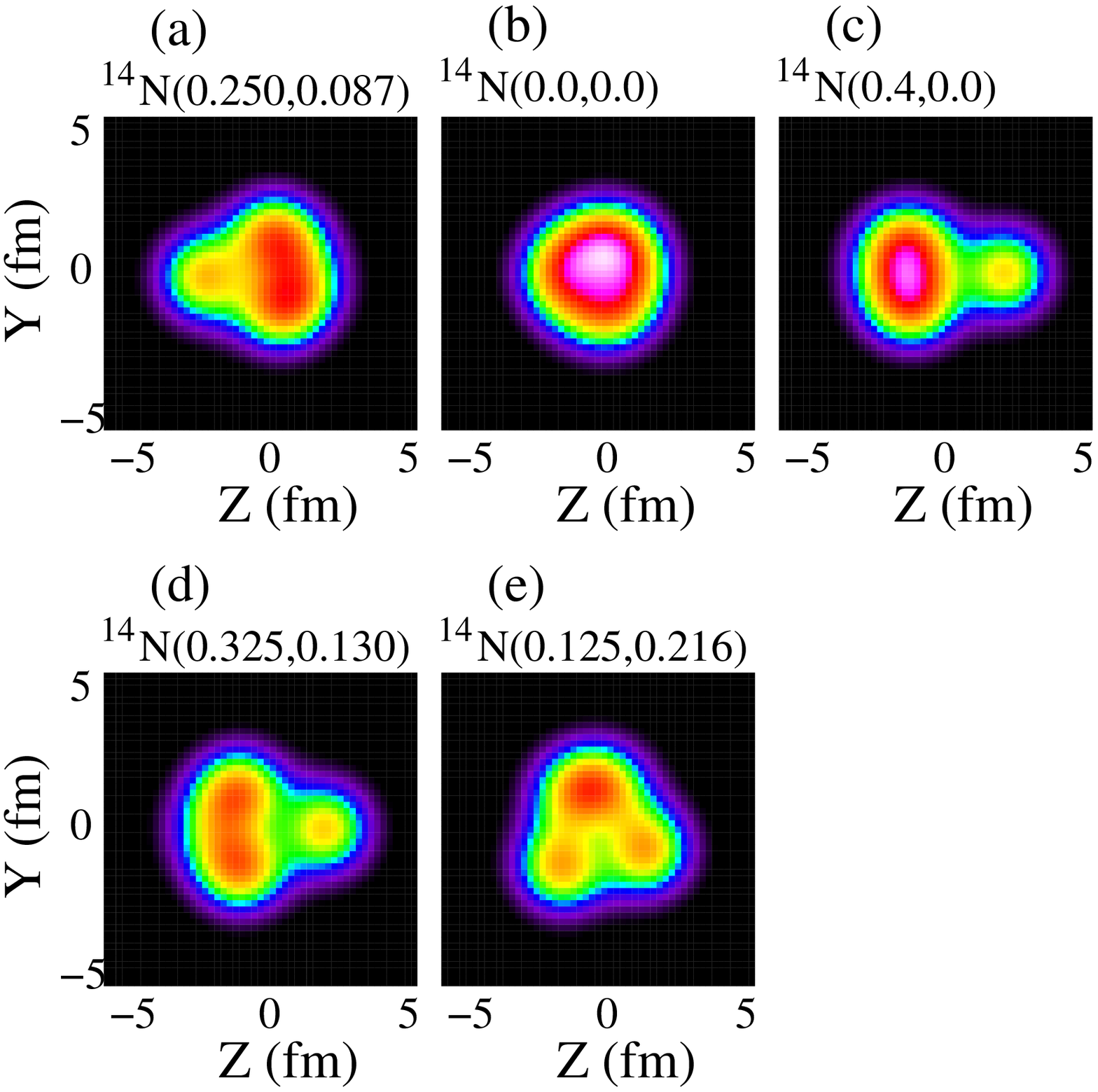}}
\caption{\label{fig:n14-bg} (Color online)
Density distribution for the intrinsic wave functions 
$\Phi_{\rm AMD}({\bf Z}^{(\beta,\gamma)})$
of $^{14}$N obtained with the $\beta$-$\gamma$ constraint AMD
using the set A interaction. 
The densities for the $J^\pi=1^+$ energy minimum 
$(\beta_{\rm min}\cos\gamma_{\rm min},
\beta_{\rm min}\sin\gamma_{\rm min})=(0.250,0.087)$ state,
and the spherical $\beta=0$, prolate 
$(\beta\cos\gamma,\beta\sin\gamma)=(0.4,0.0)$, triaxial $(0.325,0.130)$, 
and oblate $(0.125,0.216)$ states are shown.
$x$, $y$, and  $z$ axes are chosen as 
$\langle x^2 \rangle \le \langle y^2 \rangle \le \langle z^2 \rangle$,
and the matter density integrated along the
$x$ axis is plotted on the $z$-$y$ plane. 
}
\end{figure}

In Fig.~\ref{fig:c14-bg-ene}, we show
the energy expectation values for the parity projected 
wave functions
$P^\pi\Phi_{\rm AMD}({\bf Z}^{(\beta_i,\gamma_i)})$
obtained by the $\beta$-$\gamma$ constraint AMD, 
and those for the spin-parity
projected wave functions 
$P^{J\pi}_{MK}\Phi_{\rm AMD}({\bf Z}^{(\beta_i,\gamma_i)})$.
As seen in the density distributions in Figs.~\ref{fig:c14-bg} and \ref{fig:n14-bg}, 
we obtain various structures having cluster correlations 
in the $\beta$-$\gamma$ constraint AMD wave functions
of $^{14}$C and $^{14}$N. 
In the energy surface 
before the spin projection 
(Fig.~\ref{fig:c14-bg-ene}(a) and (d)), the spherical $\beta=0$ state
(Fig.~\ref{fig:c14-bg} (b) and Fig.~\ref{fig:n14-bg} (b)) 
is the energy minimum. On the 
other hand, in the $J^\pi=0^+$ projected states 
of $^{14}$C (Fig.~\ref{fig:c14-bg-ene}(b)) and the $J^\pi=1^+$ projected states 
of $^{14}$N (Fig.~\ref{fig:c14-bg-ene}(e)),
the energy minima shift to the finite $\beta$ and $\gamma$ states
at $(\beta_{\rm min}\cos\gamma_{\rm min},\beta_{\rm min}\sin\gamma_{\rm min})=(0.225,0.130)$ for $^{14}$C$(0^+)$ and 
$(\beta_{\rm min}\cos\gamma_{\rm min},\beta_{\rm min}\sin\gamma_{\rm min})=(0.250,0.087)$ for $^{14}$N$(1^+)$,
which show $\alpha$-like or $t$-like cluster correlations 
(Fig.~\ref{fig:c14-bg}(a) and Fig.~\ref{fig:n14-bg}(a)).
This indicates that the deformed states with the cluster correlations
are favored in the calculation with the spin projection,
though the spherical $0\hbar\omega$ states are favored in the model space
without the spin projection.
Moreover, on the $\beta$-$\gamma$ plane,
the $J^\pi$-projected energy surface is very soft 
over a wide area that covers 
various $\beta$-$\gamma$ states having further developed 
cluster structures. For example, in the case of 
the $J^\pi=0^+$ energy surface of $^{14}$C, 
the energy soft area within a few MeV energy difference 
from the energy minimum covers 
$(\beta\cos\gamma,\beta\sin\gamma)=(0.4,0.0)$, 
$(0.325,0.130)$, and $(0.125,0.216)$ states 
[Fig.~\ref{fig:c14-bg} (c, d, e)]
as well as the $\beta=0$ state.
These states have $\alpha$-like cluster structures
and contribute to the 
significant mixing of the excited components such as
the $2\hbar\omega$ and higher-shell 
configurations in the  
final state wave functions 
$\Psi^{^{14}{\rm C}}_{\beta\gamma\mathchar`-{\rm MC}}(J^\pi_n)$
as well as the initial state wave function
$\Psi^{^{14}{\rm N}}_{\beta\gamma\mathchar`-{\rm MC}}(J^\pi_n)$.
It should be noted that, although 
the $^{14}$C wave functions with finite $\beta$ and $\gamma$ 
show $\alpha$-like cluster structures, they do not have 
neutron and proton excitations of equal weight but do have dominant 
neutron and minor proton excitations. The reason is that 
neutrons in the $\alpha$-like cluster are more largely affected by 
the antisymmetrization with neutrons inside the $^{10}$Be core
and, therefore, are more likely excited to higher configurations
than protons.


\begin{figure}[th]
\epsfxsize=0.35\textwidth
\centerline{\epsffile{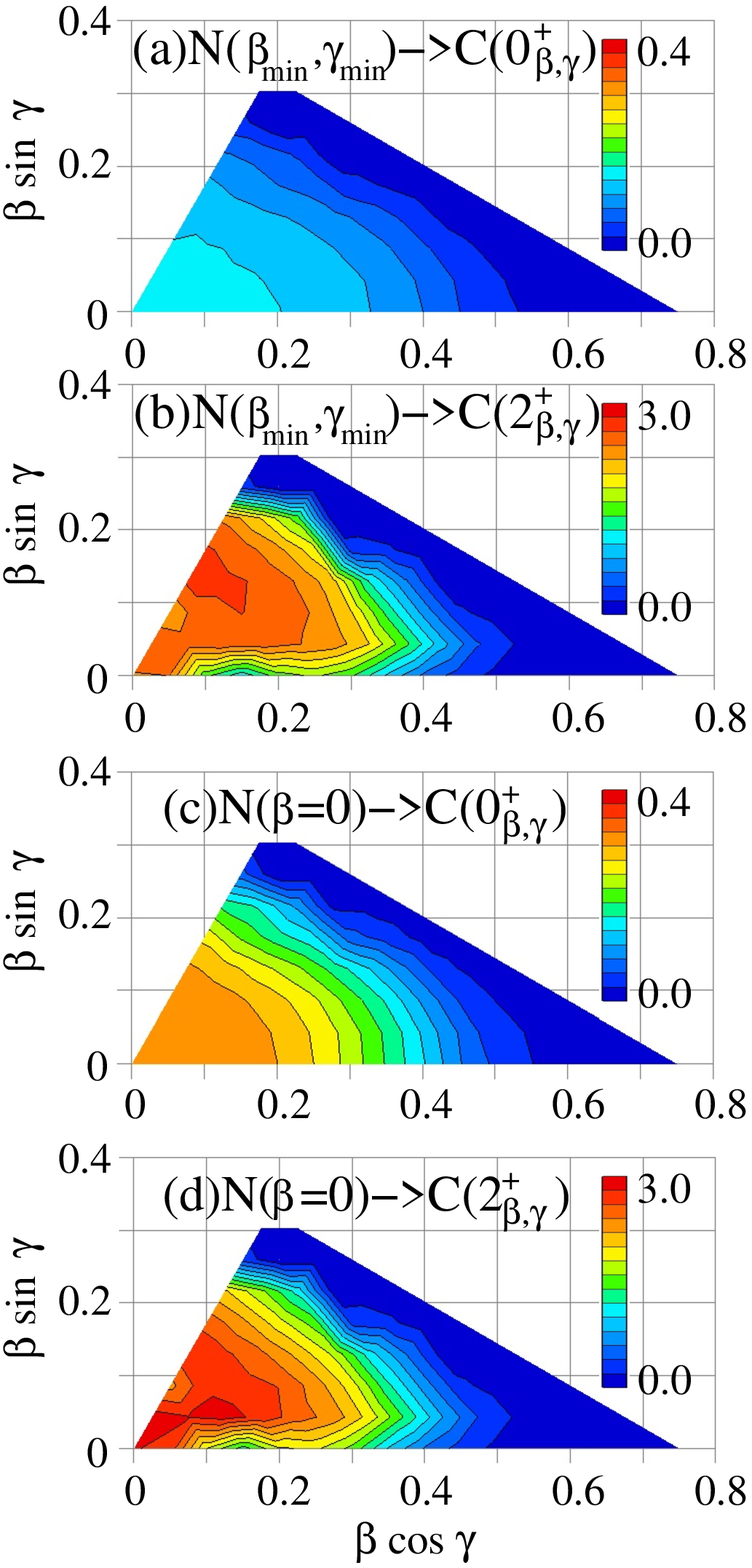}}
\caption{\label{fig:c14-bg-gt} (Color online)
$B(GT)$ values evaluated by the GT matrix element  
obtained using the single $\beta$-$\gamma$ constraint AMD 
wave function $\Phi_{\rm AMD}({\bf Z}^{(\beta,\gamma)})$ 
for $^{14}$N and that for $^{14}$C.
(a) and (b) $B(GT)$ values for the $J^\pi=0^+$ and  $2^+$ states of 
$^{14}$C projected from $\Phi_{\rm AMD}({\bf Z}^{(\beta,\gamma)})$.
The initial state is the $1^+$ energy minimum state of $^{14}$N at
$(\beta_{\rm min}\cos\gamma_{\rm min},\beta_{\rm min}\sin\gamma_{\rm min})=(0.250,0.087)$.
(c) and (d) same as (a) and (b) but the initial $^{14}$N state is 
the $1^+$ state projected from 
the spherical $\beta=0$ wave function. 
}
\end{figure}

\subsection{Effect of cluster correlation on GT strength}
As discussed above,  
the ground-state wave functions of $^{14}$C and $^{14}$N 
have cluster correlations that result in 
significant $2\hbar\omega$ and $4\hbar\omega$ components
in terms of spherical HO shell-model expansion.
In this section, we discuss the effect of the cluster correlations
on the GT strengths for $^{14}$C($0^+_1$) and $^{14}$C($2^+_1$).

In the NCSM calculation in Ref.~\cite{Aroua:2002xm}, the GT matrix elements
are sensitive to the model space of shell-model configurations. 
For example, the $B(GT)$ for the GT transition to $^{14}$C($0^+_1$)
is $B(GT)=2.518$ in the $0\hbar\omega$ model space
NCSM calculations using two-body effective interactions derived from the 
AV8' force but it is reduced to $B(GT)=0.164$ in the $6\hbar\omega$
model space calculation. The $B(GT)$ values of
the $6\hbar\omega$ NCSM calculation for 
$^{14}$C($0^+_1$) and $^{14}$C($2^+_1$) are eventually 
comparable with the present $\beta$-$\gamma$ constraint 
AMD+GCM results in spite of differences in 
the effective interactions and the model space.
Herein, we discuss the mixing effect of higher-shell components
on $B(GT)$ from the standpoint of cluster correlations.

In the $\beta$-$\gamma$ constraint AMD and the AMD+VAP calculations,
it is found that the deformed states with the cluster correlations
are favored in the calculation with the spin projection 
though the spherical $0\hbar\omega$ states are favored in the model space
without the spin projection. 
To see the effect of the cluster correlations in
finite $\beta$ and $\gamma$ states on $B(GT)$, we show 
in Fig.~\ref{fig:c14-bg-gt} the 
$B(GT)$ values evaluated by 
the GT matrix element obtained using the single $\beta$-$\gamma$ constraint AMD 
wave function for $^{14}$N and that for $^{14}$C, that is,  
the GT matrix element for the initial $1^+$ state of $^{14}$N 
projected from the $\beta$-$\gamma$ constraint AMD wave function
at the $1^+$-projected energy minimum 
$(\beta_{\rm min}\cos\gamma_{\rm min},\beta_{\rm min}\sin\gamma_{\rm min})=(0.250,0.087)$ and the final $J^\pi=0^+$ and $2^+$ states 
of $^{14}$C projected from the
$\beta$-$\gamma$ constraint AMD wave function 
$\Phi_{\rm AMD}({\bf Z}^{(\beta,\gamma)})$.
We also show the $B(GT)$ given by the GT matrix element
for the case in which the initial $^{14}$N state is the 
spherical $\beta=0$ state.
Here the $K$-mixing is taken into account.
In both cases of initial $^{14}$N states, the $\beta_{\rm min}$-$\gamma_{\rm min}$ and $\beta=0$ states, 
the $B(GT)$ for $^{14}$C$(0^+)$ decreases 
as the deformation $\beta$ of $^{14}$C increases. 
Comparing the $B(GT)$ value for the final 
$^{14}$C state at the spherical limit $\beta=0$
with that for the $^{14}$C state at the $J^\pi=0^+$ energy minimum 
$(\beta_{\rm min}\cos\gamma_{\rm min},
\beta_{\rm min}\sin\gamma_{\rm min})=(0.225,0.130)$, 
the $B(GT)$ to 
$^{14}$C$(0^+)$ is reduced by 30\% because of the 
cluster correlation in the final $^{14}$C state.
In the comparison of the $B(GT)$ for the initial $^{14}$N state 
with  $\beta=0$  and $\beta_{\rm min}$-$\gamma_{\rm min}$, a
50\% reduction of $B(GT)$
occurs because of the cluster correlation in the initial $^{14}$N state.
In contrast to the $B(GT)$ for $^{14}$C$(0^+)$, no reduction caused
by cluster correlation is seen 
in the $B(GT)$ for $^{14}$C$(2^+)$.

As mentioned before, the present results overestimate the 
experimental $B(GT)$ for $^{14}$C$(0^+_1)$ and $^{14}$C$(2^+_1)$.
Although the $B(GT)$ for $^{14}$C$(0^+_1)$ can 
be somewhat reduced by the cluster correlation, the possible 
reduction is only a factor 2$-$3 at most, and it is difficult to 
describe the anomalous suppression of the GT matrix element
known from the long life time of $^{14}$C.
For the transition to $^{14}$C$(2^+_1)$, 
the $B(GT)$ seems insensitive to 
the cluster correlation, and it is also difficult to quantitatively reproduce 
the experimental data in the present calculation. 

\section{Summary}\label{sec:summary}

GT transitions from the $^{14}$N ground state to the 
$^{14}$C ground and excited states were investigated on the basis of 
the model of AMD.
The AMD+VAP method and the $\beta$-$\gamma$ constraint AMD+GCM 
were applied to $0^+$, $1^+$, and $2^+$ states of $^{14}$C
as well as the ground state of $^{14}$N. Both calculations 
show similar results.

The calculated strengths for the allowed transitions to 
$0^+$, $1^+$, and $2^+$ states of $^{14}$C were compared with 
experimental data measured by high-resolution
charge exchange reactions.
The calculated $GT$ transition to the $2^+_1$ state is strong, whereas 
those to 
the $0^+_{2,3}$ and $2^+_{2,3}$ states having 
dominant $2\hbar\omega$ excited configurations are relatively weak.
The $B(GT)$ distributions to excited states of $^{14}$C in the present
calculations are in reasonable agreement with the experimental 
$B(GT)$ distributions except for $B(GT)$ for $^{14}{\rm C}(2^+_1)$. 
The present calculation can not describe the anonymously long life time of $^{14}$C, though the GT strength of the $^{14}$C ground state 
is relatively small compared with the $2^+_{1,2}$ and $1^+_1$ states.

Compared with the large-scale NCSM calculations \cite{Aroua:2002xm}, 
the $B(GT)$ values for $^{14}{\rm C}(0^+_1)$ and $^{14}{\rm C}(2^+_1)$ in the present calculation are almost the same as those in the 
$6\hbar\omega$ NCSM calculations. For higher $0^+$ and $2^+$ states, 
the present calculation shows a better description 
of the experimental $B(GT)$ distributions 
in the $E_x\sim 10-15$ MeV region. 

It was found that the ground-state wave functions 
of $^{14}$C and $^{14}$N 
have cluster correlations that result in 
significant $2\hbar\omega$ and $4\hbar\omega$ components
in terms of spherical HO shell-model expansion.
In the excited states of $^{14}$C, 
further development of cluster structures is seen.

The effect of the cluster correlations
on the GT strengths for $^{14}$C($0^+_1$) and $^{14}$C($2^+_1$)
was discussed.
Although the $B(GT)$ for $^{14}$C$(0^+_1)$ can 
be somewhat reduced by the cluster correlation, the possible 
reduction is only a factor of 2$-$3 at most, and it is difficult to 
describe the anomalous suppression of the GT matrix element
known from the long life time of $^{14}$C.

\section*{Acknowledgments} 
The authors would like to thank Prof. Fujita for valuable discussions.
The computational calculations in this study were performed on the
supercomputers at YITP, Kyoto University.
This work was supported by a Grant-in-Aid for Scientific Research from the Japan Society for the Promotion of Science (JSPS).
It was also supported by 
a Grant-in-Aid for the Global COE Program, "The Next Generation of Physics, 
Spun from Universality and Emergence", from the Ministry of Education, Culture, Sports, Science and Technology (MEXT) of Japan.

\end{document}